\documentclass{emulateapj} 

\def\lsim{~\raise0.3ex\hbox{$<$}\kern-0.75em{\lower0.65ex\hbox{$\sim$}}~}
\def\gsim{~\raise0.3ex\hbox{$>$}\kern-0.75em{\lower0.65ex\hbox{$\sim$}}~}
\def\lt{~\hbox{$<$}~}

\def\lbrack2{[\![}
\def\rbrack2{]\!]}

\def\vcirc{{V_{\rm c}}}

\def\kms{{\rm\,km\,s^{-1}}}
\def\msunyr{{\rm\,M_\odot\,s^{-1}}}
\def\kpc{{\rm\,kpc}}
\def\mpc{{\rm\,Mpc}}

\def\ev{{\rm\,eV}}
\def\msun{{\rm\,M_\odot}}

\def\pc{{\rm\,pc}}

\def\Myrs{{\rm\,Myrs}}

\def\fesc{{f_{\rm esc}}}

\shorttitle{Modeling LyC emission from young galaxies}
\shortauthors{Razoumov \& Sommer-Larsen}

\begin{document}


\title{Modeling Lyman continuum emission from young galaxies}


\author{Alexei O. Razoumov\altaffilmark{1}}
\email{razoumov@ap.smu.ca}

\author{Jesper Sommer-Larsen\altaffilmark{2,3}}
\email{jslarsen@dark-cosmology.dk}


\altaffiltext{1}{Institute for Computational Astrophysics, Dept. of
  Astronomy \& Physics, Saint Mary's University, Halifax, NS, B3H 3C3,
  Canada}

\altaffiltext{2}{Dark Cosmology Centre, Niels Bohr Institute,
  University of Copenhagen, Juliane Maries Vej 30, DK-2100 Copenhagen,
  Denmark}

\altaffiltext{3}{Institute of Astronomy, University of Tokyo, Osawa
  2-21-1, Mitaka, Tokyo, 181-0015, Japan}


\begin{abstract}
  Based on cosmological simulations, we model Lyman continuum emission
  from a sample of 11 high-redshift star forming galaxies spanning a
  mass range of a factor 20. Each of the 11 galaxies has been
  simulated both with a Salpeter and a Kroupa initial mass function
  (IMF). We find that the Lyman continuum (LyC) luminosity of an
  average star forming galaxy in our sample declines from $z=3.6$ to
  2.4 due to the steady gas infall and higher gas clumping at lower
  redshifts, increasingly hampering the escape of ionizing radiation.
  The galaxy-to-galaxy variation of apparent LyC emission at a fixed
  redshift is caused in approximately equal parts by the intrinsic
  variations in the LyC emission and by orientation effects. The
  combined scatter of an order of magnitude can explain the variance
  in the far-UV spectra of high-redshift galaxies detected by
  \citet{shapley....06}. Our results imply that the cosmic galactic
  ionizing UV luminosity would be monotonically decreasing from
  $z=3.6$ to 2.4, curiously anti-correlated with the star formation
  rate in the smaller galaxies, which on average rises during this
  redshift interval.
\end{abstract}



\section{Introduction}

Recent spectroscopic detection of Lyman continuum (LyC) emission from
individual $z\sim3$ galaxies \citep{shapley....06} represents a major
step forward in analyzing the stellar contribution to the ionizing
background at high redshifts. In a study of 14 $z\sim3$ star-forming
(SF) galaxies \citet{shapley....06} have detected escaping ionizing
radiation from two objects, concluding that there is significant
variance in the emergent LyC spectrum. The nature of this variance is
currently not understood, although it is undoubtedly linked to the
physical conditions in the interstellar medium (ISM) at these high
redshifts. Comparison of observed LyC emission to theoretical models
should in principle allow us to put constraints on the physics of SF
in young galaxies and study the effect of stellar radiation on the
thermal properties of high-redshift gas, including its role in
maintaining cosmic reionization. One of the main parameters in this
study is the escape fraction of ionizing photons from the clumpy ISM,
a quantity of primary importance in determining the contribution of
the volumetric stellar luminosity to the ionizing background.

In \citet[][Paper I]{razoumov.06} we presented calculations of the
escape fraction of LyC photons from a large number of SF regions in
two simulated high-redshift proto-galaxies. We found that the modeled
redshift evolution of $\fesc$ matches the observational findings of
\citet{inoue..06}, namely the decline from $\fesc\sim6-10\%$ at
$z\gsim3.6$ to $\fesc\sim1-2\%$ at $z=2.4$. This decline is attributed
to a much higher clustering of gas around the SF regions at lower
redshifts, due to accretion of the intergalactic gas onto growing
proto-galaxies. In this paper we extend these calculations to a larger
set of (proto-) spiral galaxies covering a range of masses and study
the dependence of our results on the amount of feedback per unit
stellar mass encoded in the initial mass function (IMF). We moreover
determine galaxy-to-galaxy LyC emission variations at a given redshift
and galaxy mass, as well as variations for a given galaxy along
different lines of sight.

\section{Models}

We use results of high-resolution galaxy formation simulations in a
standard LCDM cosmology done with a significantly improved version of
the TreeSPH code described by \citet{sommer-larsen..03}. The code
employs the ``conservative'' entropy formulation \citep{springel.02},
non-instantaneous gas recycling and chemical evolution with 10
elements \citep{lia..2002}, metallicity-dependent atomic radiative
cooling, and self-shielding in regions with the mean free path of
Lyman limit photons below $1\kpc$. All galaxies for this project were
selected from a dark matter-only run of box length $10h^{-1}\mpc$,
which was then resimulated with the TreeSPH code at higher resolution
in Lagrangian regions enclosing the galaxies. Further details on this
simulation set can be found in
\citet{sommer-larsen06}. Table~\ref{models} lists 11 galaxies, in
order of increasing mass, for which we performed radiative transfer
calculations around all their SF regions at $z=3.6$, 2.95 and
2.39. The first five are fairly small ($\vcirc\lsim130\kms$, see
below), gas-rich galaxies that exhibit relatively little SF. Galaxy 41
is an intermediate-size system with $\vcirc=150\kms$, whereas the rest
are larger ($\vcirc\gsim180\kms$) disk galaxies. The most massive
galaxies were computed at ``normal'' resolution, with SPH (and star)
particle masses of $1.1\times10^6\msun$ and interpolated grid
resolution of $30\pc$, the smaller galaxies were simulated with 8
times higher mass resolution and twice the force resolution. The only
``low'' resolution simulation was the $\vcirc=310\kms$ galaxy with
particle masses of $3.1\times10^6\msun$.

All galaxies in Table~\ref{models} are labeled by their $z=0$
characteristic circular velocities, which were calculated using the
technique described in \citet{sommer-larsen.01}. For our purpose, the
present-day $\vcirc$ is a good measure of $z\sim3$ galaxy dynamical
masses for a number of reasons. First, galaxies tend to grow in size
from $z=3$ to $z=0$, but not so much in $\vcirc$. In addition, there
is a fairly tight relation between the $z=3$ stellar and virial masses
and the $z=0$ ones. This is simply explained by the fact that our
sample contains only field galaxies, which become disks by $z=0$. All
of them have relatively quiet $z<3$ merging histories; in fact, at
$z=3$ it is already fairly easy to identify the main
proto-galaxy. Finally, most of our $z=3$ systems have several
proto-galactic components, and our results at $r=100\kpc$ account for
absorption and stellar emission in all of these components, not just
the main proto-galaxy. Note that stellar masses, on the other hand,
are sensitive to the feedback scenario as we will show in Section 3.1,
and cannot be uniquely described by $\vcirc$ which is rather the
measure of the galaxy's dynamical mass.

We model SF with a set of discrete star ``particles'', which represent
a population of stars born at almost the same time in accordance with
a given IMF. The stellar UV luminosity is determined using the
population synthesis package Starburst 1999
\citep{leitherer........99} with continuous SF distributed among all
stars younger than $34\Myrs$. To compute the time-dependent ionization
by stellar photons, we use the point source radiative transfer
algorithm on adaptively refined meshes first employed in Paper I. This
algorithm extends the adaptive ray-splitting scheme of \citet{abel.02}
to a model with variable grid resolution. Around each SF region we
build a system of $12\times 4^{n-1}$ radial rays that split either as
we move farther away from the source or as we enter a refined
cell. Once a radial ray is refined angularly, it stays refined at
larger distances from the source, even if we leave the high-resolution
region. In each cell we accumulate the ionization and heating rates
due to photons travelling along ray segments passing through that
cell. These rates are then used to update temperature and the
ionization state of hydrogen and helium, as a function of time.

To study the sensitivity of our results to the strength of stellar
feedback which is in turn a function of the number of massive stars in
our SF regions, we adopt two different but widely used IMFs, the
standard \citet{salpeter55} and the triple-interval \citet{kroupa98}
IMF. The Salpeter IMF produces approximately twice as much SNII
feedback and stellar ionizing radiation per unit stellar mass compared
to the Kroupa IMF. However, both IMFs result in less feedback than the
more top-heavy IMF used in Paper I. The SNII feedback and chemical
enrichment are included into the hydrodynamical models, whereas the
propagation of ionizing stellar photons is traced with post-process
radiative transfer, as detailed in Paper I.

We do not include the effects of dust in our current calculations. The
fact that our models in Paper I which did not include dust matched
well the observationally determined escape fractions of
\citet{inoue..06} as a function of z, for $z\sim 2.4-3.6$, can be
taken as indirect evidence that dust may be not of primary importance
in controlling LyC escape fractions and luminosities of young
galaxies. On the other hand, a number of physical processes such as
cooling and molecule formation depend on dust density and
composition. Dust forms from heavier elements, and metallicity of our
galaxies is fairly high, e.g. the iron abundances range from $[\rm
Fe/H]\sim-1.2$ for the smallest galaxies to $[\rm Fe/H]\sim -0.4$ for
the largest. Unfortunately, the physics of dust grain growth and
destruction is extremely complicated and is well beyond the scope of
the present paper, although we are planning to include a simple
parametric description into our future simulations.

\begin{table}
  \begin{center}
    \caption{Galaxy pairs Salpeter -- Kroupa.\label{models}}
    \begin{tabular}{cccc}
      \tableline
      Galaxy & resolution & $\vcirc$(z=0) & comment\\
      && $\kms$ &\\
      \tableline
      84 & high & 115 & small galaxy\\
      93 & high & 122 & ''\\
      115 & high & 125 & ''\\
      108 & high & 131 & ''\\
      87 & high & 132 & ''\\
      41 & normal & 150 & intermediate-size\\
      33 & normal & 180 & sub-Milky Way\\
      29 & normal & 205 & slightly sub-Milky Way\\
      26 & normal & 208 & ''\\
      15 & normal & 245 & M31-like disk galaxy\\
      15\_sc1.39 & low & 310 & very large disk galaxy\\
      \tableline
    \end{tabular}
  \end{center}
\end{table}
\section{Results}


\subsection{Redshift evolution and the effect of the IMF}

Following Paper I, we define $\fesc$ simply as the fraction of
(ionizing) stellar photons of a given frequency that reach distance
$r$ from the SF regions. Here all our results are computed at
$r=100\kpc$, i.e. on a sphere that encompasses all absorbing clouds
associated with each galaxy, at $t=10\Myrs$ after the stellar sources
were switched on. However, to obtain these results, the full 3D
time-dependent ionization structure of the gas was computed.

\begin{figure}
  \epsscale{1.1}\plotone{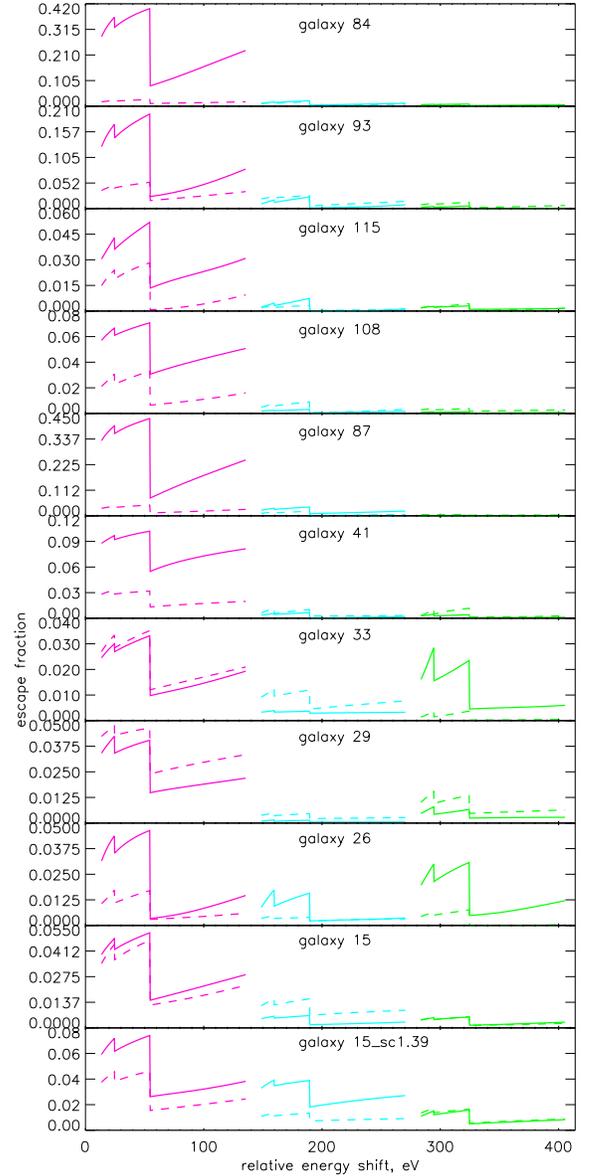}
  \caption{Spectral dependence of $\fesc$ at $r=100\kpc$ for all
    Salpeter (solid lines) -- Kroupa (dashed lines) galaxy pairs at
    $z=3.6$ (magenta), $z=2.95$ (cyan), and $z=2.39$ (green). For each
    model the redshift evolution is from left to right, and each curve
    goes from $13.6\ev$ to $135\ev$. Galaxies are listed in order of
    increasing mass (top to bottom).}
  \label{escapeFraction-K98-S}
\end{figure}

\begin{figure}
  \epsscale{1.1}\plotone{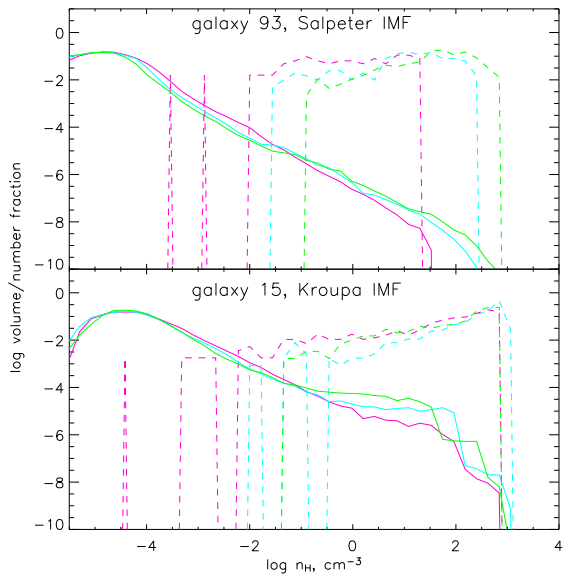}
  \caption{PDF of gas (solid lines) and stars (dashed lines) in
    galaxies 93 (top, evolved with Salpeter IMF) and 15 (bottom,
    evolved with Kroupa IMF) at $z=3.6$ (magenta), $z=2.95$ (cyan),
    and $z=2.39$ (green). The PDF is the fraction of total volume for
    gas and the fraction of total number for stars in a given density
    interval $\Delta\log\rho$.}
  \label{pdf-S93-06-64-K15-06-8}
\end{figure}

In Fig.~\ref{escapeFraction-K98-S} we plot the angle- and
source-averaged escape fractions for all galaxies from
Table~\ref{models} as a function of the photon energy, at $z=3.6$,
2.95 and 2.39. We confirm our earlier conclusion \citep{razoumov.06}
that $\fesc$ shows tendency to decrease with time due to increased
clustering of gas near the SF regions. This can be illustrated by
plotting the density distribution of gas and stars in individual
galaxies as a function of redshift. Fig.~\ref{pdf-S93-06-64-K15-06-8}
shows the probability distribution function (PDF) of gas density and
the fraction of stars hosted by cells of that density for two typical
systems in our simulations: a small galaxy 93 ($\vcirc=122\kms$) with
the Salpeter IMF and a much larger galaxy 15 ($\vcirc=245\kms$) with
the Kroupa IMF. Higher gas clustering near SF regions at $z<3$ is
evident. Note that the horizontal axis in this plot does not
necessarily indicate the densities at which stars form but rather
densities at which star particles are found at a specific
redshift. Feedback tends to evacuate gas from the SF regions,
therefore, some stellar particles can be hosted by cells with very
little gas left.

\begin{figure}
  \epsscale{1.1}\plotone{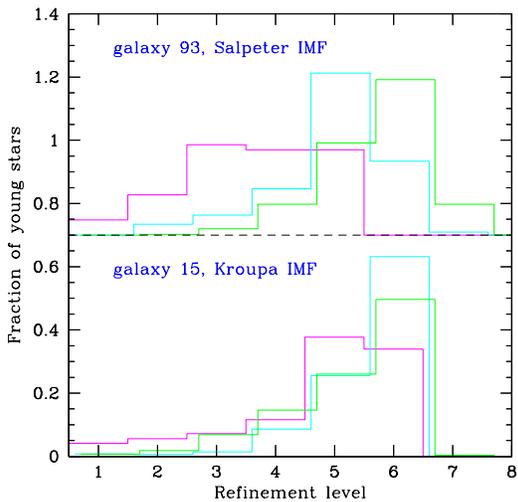}
  \caption{The fraction of young stars as a function of refinement
    level for the two galaxies from Fig.~\ref{pdf-S93-06-64-K15-06-8}
    at $z=3.6$ (magenta), $z=2.95$ (cyan), and $z=2.39$ (green). The
    top plot has been displaced vertically for clarity.}
  \label{levels-S93-06-64-K15-06-8}
\end{figure}

Fig.~\ref{levels-S93-06-64-K15-06-8} shows on a linear scale the
fraction of stellar particles younger than $34\Myrs$, as a function of
the refinement level, for the same two galaxies. The nested grid
structure was created from the SPH particle distribution using the gas
density as a refinement criterion. Stars form in highly overdense
regions the average density of which increases with time as more gas
accretes onto growing proto-galaxies from the intergalactic
medium. This is particularly evident in small galaxies such as 93,
with a gradual rise in the average density of SF regions from $z=3.6$
to $z=2.39$, leading to a decline in ionizing UV output. Note that all
lower mass galaxies in Fig.~\ref{escapeFraction-K98-S} display a drop
in $\fesc$ in this redshift interval. Larger field galaxies such as 15
already accumulated a massive amount of gas by $z=3.6$ and exhibit
somewhat more limited growth at lower redshifts, while their SF
regions are skewed to more dense environments. Consequently, massive
galaxies have already fairly small $\fesc$ even at early times, and in
some cases they can even show a rise in $\fesc$ at lower redshifts,
such as in galaxies 33, 29, 26 with the Salpeter IMF. Overall, we find
significant variations from galaxy to galaxy, especially in the
lower-mass systems, with the Salpeter IMF Lyman-limit $\fesc$ varying
from $3\%$ in 115 to $33\%$ in 87.

\citet{sommer-larsen..03} found that early, fairly strong bursts of SF
converting several percent of the initial gas mass into stars with the
Salpeter IMF can substantially alleviate the disk galaxy angular
momentum problem, as feedback from these starbursts can blow a larger
fraction of the remaining gas out of the small proto-galactic clumps,
with this gas later gradually settling to form extended disks. These
models were later expanded to include the Kroupa IMF
\citep{sommer-larsen06}, and at the moment they still underpredict the
observed angular momenta of $z=0$ disk galaxies by approximately a
factor of $1.5$. On the other hand, these models feature fairly
realistic high-redshift systems which reproduce the sub-mm properties
of Lyman break galaxies \citep{greve.06}, show extended Ly$\alpha$
emission \citep{laursen.07}, and have fairly small stellar cores at
$z=3$ consistent with observations \citep{trujillo.06,dahlen......07}.

\begin{figure}
  \epsscale{1.1}\plotone{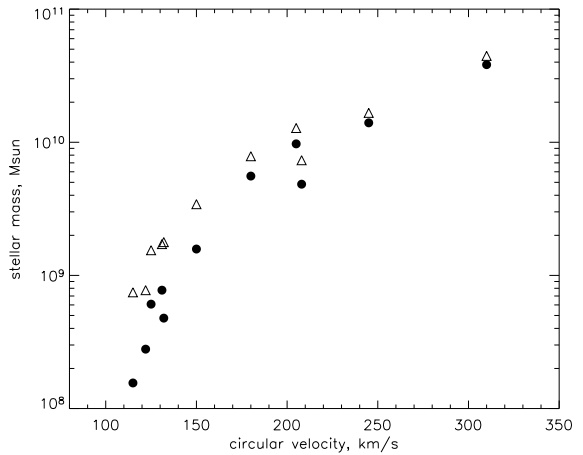}
  \caption{Total $z=3$ stellar mass vs. the characteristic circular
    velocity for Salpeter (filled circles) and Kroupa (open triangles)
    galaxies.}
  \label{stellarMass}
\end{figure}

\begin{figure}
  \epsscale{1.1}\plotone{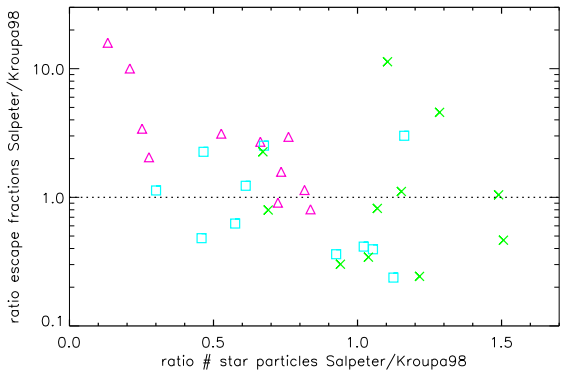}
  \caption{Ratio $f_{\rm esc,S}/f_{\rm esc,K}$ of the Lyman-limit
    escape fractions for models with the Salpeter and Kroupa IMFs
    vs. the ratio $N_{*,\rm S}/N_{*,\rm K}$ of the number of young
    ($t_{\rm age}<34\Myrs$) stellar particles in each pair of models,
    at $z=3.6$ (magenta), $z=2.95$ (cyan), and $z=2.39$ (green).}
  \label{doubleRatio-K98-S}
\end{figure}

\begin{figure}
  \epsscale{1.1}\plotone{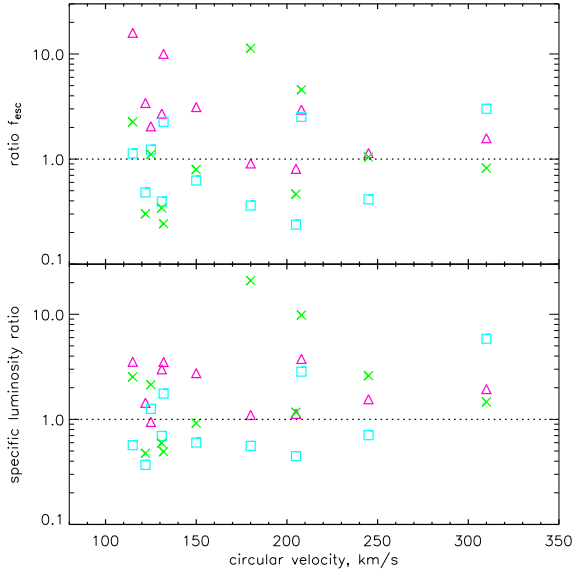}
  \caption{Ratio $f_{\rm esc,S}/f_{\rm esc,K}$ of the Lyman-limit
    escape fractions (top) and the Lyman-limit specific luminosities
    (bottom) for models with the Salpeter and Kroupa IMFs vs. the
    characteristic circular velocity, at $z=3.6$ (magenta), $z=2.95$
    (cyan), and $z=2.39$ (green).}
  \label{ratio-K98-S}
\end{figure}

We now focus on the effect of the IMF on LyC emission. With the
Salpeter IMF the stronger feedback at early times suppresses SF
relative to the Kroupa case yielding less massive $z\sim3$ stellar
components (Fig.~\ref{stellarMass}), and building up a larger
reservoir of hot gas around the galaxy. This results in an
anti-correlation between $\fesc$ and the number of young ($t_{\rm
  age}<34\Myrs$) stars at $z=3.6$ (Fig.~\ref{doubleRatio-K98-S}): the
stronger feedback there is in a galaxy, the fewer stars it forms, but
also the higher $\fesc$ it gets as the same amount of gas is being
dispersed over a larger volume, and more neutral gas is ionized per
unit stellar mass. This anti-correlation reduces the sensitivity of
the absolute UV luminosity to the fraction of massive stars encoded in
the IMF. For example, in galaxy 87 ($v_c = 132\kms$) at $z=3.6$ its
Salpeter $\fesc$ is about 10 higher than the Kroupa $\fesc$
(Fig.~\ref{ratio-K98-S}a), however, it has 93 young stellar particles
with the Salpeter IMF, while the Kroupa IMF results in 444
particles. The result is that its Salpeter luminosity is only $\sim3$
times higher than its Kroupa luminosity (Fig.~\ref{ratio-K98-S}b).

Eventually, the feedback cannot prevent the cool-out of hot gas, which
sooner or later leads to a rise in SF. In fact, Salpeter galaxies tend
to produce more stars than Kroupa galaxies at later times, as can be
seen in the number of young stars along the horizontal axis in
Fig.~\ref{doubleRatio-K98-S} at $z=2.39$. By this time, however, there
is no correlation between the escape fractions and the number of
recent starbursts, as consistently stronger feedback has a very
non-linear effect on the gas distribution at lower redshift. On one
hand, models with higher feedback rates per unit stellar mass
generally have more gas left over for SF from earlier times, and by
$z=2.39$ some of this gas might have cooled back into dense clouds in
the disk ISM. Therefore, combined supernova winds now have to push
through a thicker layer of material, unless all of this cool-out gas
has been efficiently converted into stars on a short timescale which
is unlikely. On the other hand, irrespective of the earlier SF
history, stronger feedback might still clear channels through which
ionizing photons escape into the intergalactic medium. The end result
is that all these processes create a very complex porous ISM, with
lower average $\fesc$, but higher variations from galaxy to galaxy,
and a larger scatter when we vary the IMF (Fig.~\ref{ratio-K98-S}a,b),
as by now many effects contribute to the value of $\fesc$.

Gas blowout is expected to be less efficient in already formed disk
galaxies as it drives a typically much smaller fraction of
interstellar gas in bipolar outflows perpendicular to the disk
\citep{maclow.99}. As one would expect, in our models the escape
fractions show a weaker dependence on the feedback strength in the most
massive galaxies (Fig.~\ref{ratio-K98-S}a).

It is interesting that in two galaxies (29 and 33) stronger feedback
leads to lower $\fesc$ even at $z=3.6$. Close examination reveals that
with the Salpeter IMF these fairly massive galaxies expelled a large
amount of gas prior to this epoch, whereas in the Kroupa IMF case a
noticeable fraction of this gas was converted into stars early into
the run. With the Salpeter IMF, by $z=3.6$ this gas is falling back
onto the galaxies supplying cold material to the SF regions. In
addition, galaxy 29 has two components very close to each other, and
some of the expelled gas is actually trapped in the common
gravitational field before accreting back onto both components.

\begin{figure}
  \epsscale{1.1}\plotone{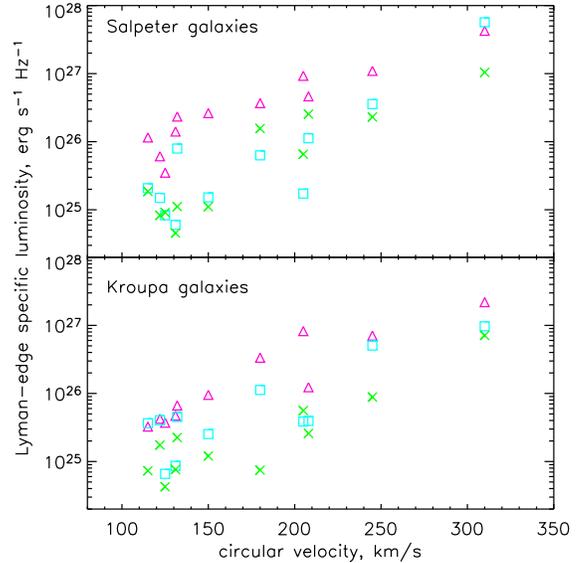}
  \caption{Lyman-limit specific luminosity vs. the characteristic
    circular velocity, at $z=3.6$ (magenta), $z=2.95$ (cyan), and
    $z=2.39$ (green), for Salpeter (top) and Kroupa (bottom)
    galaxies.}
  \label{luminosities-K98-S}
\end{figure}

The luminosity of each galaxy is a product of the original IMF
spectrum, the number of SF regions in the galaxy, the SF rate of each
region which depends on the mass of a stellar particle, and the escape
fraction. In Fig.~\ref{luminosities-K98-S} we plot the specific (per
unit frequency) Lyman-limit luminosity $L_{912}$ of each galaxy
vs. its circular velocity, at all three redshifts. Overall, the
Lyman-limit luminosities of galaxies simulated using the two different
IMFs follow similar trends with $\vcirc$ and $z$. Note that for most
systems $L_{912}$ is higher at earlier times, meaning that galaxies
become an increasingly more important source of ionizing photons
relative to quasars the number density of which declines rapidly at
$z\gsim3$ \citep{richards-06}. Therefore, the results shown in
Fig.~\ref{luminosities-K98-S} are of great importance in relation to
the ionization history of the Universe, and we shall return to this
topic in a forthcoming paper.

We also find that the average cosmic galactic LyC luminosity which in
our models decreases monotonically from $z=3.6$ to 2.4, most notably
in the low mass galaxies, is anti-correlated with the star formation
rate in the same galaxies, which rises during this redshift
interval. For the 5 small ($\vcirc\lt150\kms$) galaxies in our sample,
the average SF rates are 1.8, 2.6 and 2.7 $\msunyr$ at $z=3.6$, 2.95
and 2.39 for the Kroupa galaxies, and 0.6, 1.4 and 2.8 $\msunyr$ for
the Salpeter galaxies, respectively.  For the larger galaxies, the
evolution of the SF rate with redshift is more flat in the $z=3.6-2.4$
interval. Both above findings are in qualitative agreement with the
observational result of \citet{sawicki.06}, that the galaxy
(non-ionizing) UV luminosity function gradually rises at the faint end
from $z=4$ to $z=2$ , while the bright end of the luminosity function
exhibits virtually no evolution.

\begin{figure}
  \epsscale{1.1}\plotone{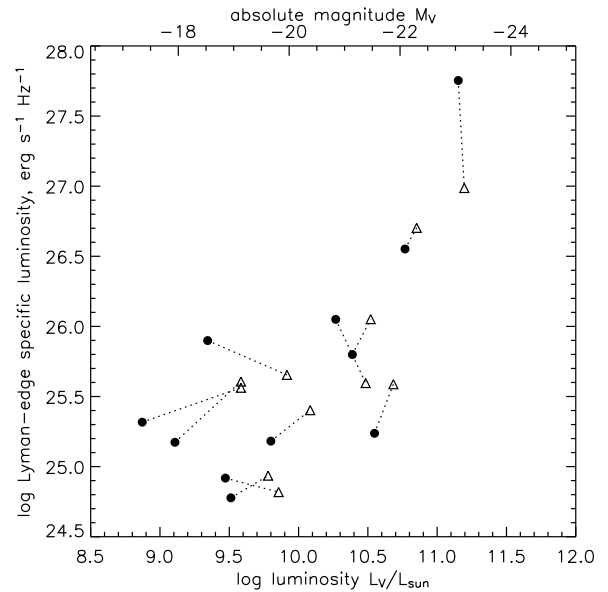}
  \caption{Lyman-limit specific luminosity at $z=2.95$ vs. the optical
    luminosity. The dotted lines connect a pair of models of each
    galaxy evolved with the Salpeter (filled circles) and Kroupa (open
    triangles) IMFs, respectively. The logarithmic scale is the same
    on the vertical and horizontal axes.}
  \label{magnitudeLuminosity}
\end{figure}

In Fig.~\ref{magnitudeLuminosity} we plot the Lyman-limit luminosities
of all galaxies in our sample vs. their optical luminosities, at
$z=2.95$. The optical luminosity is tightly coupled to the number of
bright stars, whereas the ionizing luminosity also depends on the
amount of absorbing neutral gas in the vicinity of SF regions. As gas
blowout is less efficient in more massive galaxies, the supply of
material available for SF does not depend so strongly on the feedback
strength, and the optical luminosity of such galaxies is less affected
when varying the IMF. The ionizing luminosities, on the other hand,
are sensitive to the conditions near SF regions and show variations
even in already formed massive systems. It is also seen from
Fig.~\ref{magnitudeLuminosity} that the effects of changing the IMF on
the LyC and optical luminosities, respectively, are essentially
uncorrelated.

\subsection{Radial dependence of the escape fraction}

\begin{figure}
  \epsscale{1.1}\plotone{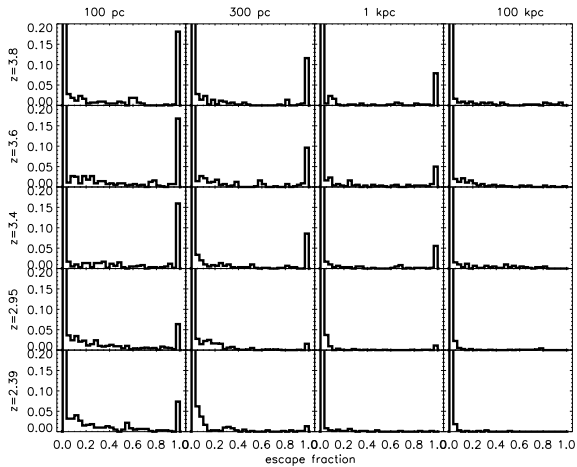}
  \caption{Distribution of sources by their Lyman-limit escape
    fractions in galaxy 15 with the Kroupa IMF at $z=3.8$, 3.6, 3.4,
    2.95 and 2.39 (top to bottom), at four different radii, from 100
    pc to 100 kpc (left to right). The vertical axis has been
    truncated to showcase the radial distribution.}
  \label{radial}
\end{figure}

To probe conditions inside the ISM, it is instructive to look at the
radial dependence of absorption as a function of redshift in one of
our galaxies. In Fig.~\ref{radial} we show the $4\pi$-averaged
distribution of $\fesc$ for all star particles in the more massive
galaxy 15, computed with the Kroupa IMF, as a function of redshift and
the distance from these particles. At all five redshifts a significant
fraction of photons reaches the radius $r=100\pc$, although this
fraction decreases with time as more gas is clustered around the SF
regions. At $r=1\kpc$ the redshift evolution is more noticeable, as
very few sources have $\fesc>0.1$ at $z=2.39$.  In other words, a
random observer sitting inside this galaxy would see LyC directly from
$15-20\%$ of the young stars in that galaxy at $z=3.8$ and from only a
few $\%$ of the stars at $z=2.39$. For a distant observer beyond this
galaxy's virial radius of $r\sim45\kpc$, the average Lyman-limit
escape fraction drops from $\sim4\%$ to $\sim0.4\%$ in the above
redshift interval.

\subsection{Orientation effects}

In Section 3.1 we saw that for galaxies of a given mass the
galaxy-to-galaxy scatter in the LyC emission can be partially
attributed to variations in the physical conditions in the clumpy ISM
{\it surrounding} the SF regions (factor of $\sim2-3$ for either of
the selected IMFs in Fig.~\ref{luminosities-K98-S}).

\begin{figure}
  \epsscale{1.}\plotone{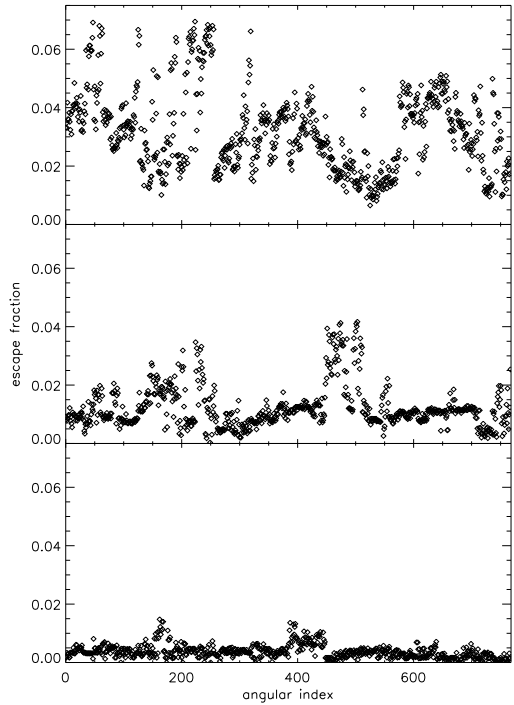}
  \caption{Angular variation of the Lyman-limit escape fraction for
    K15, at $z=3.6$, 2.95, and 2.39 (top to bottom).}
  \label{angularVariation.K15-06-8}
\end{figure}

\begin{figure}
  \epsscale{1.}\plotone{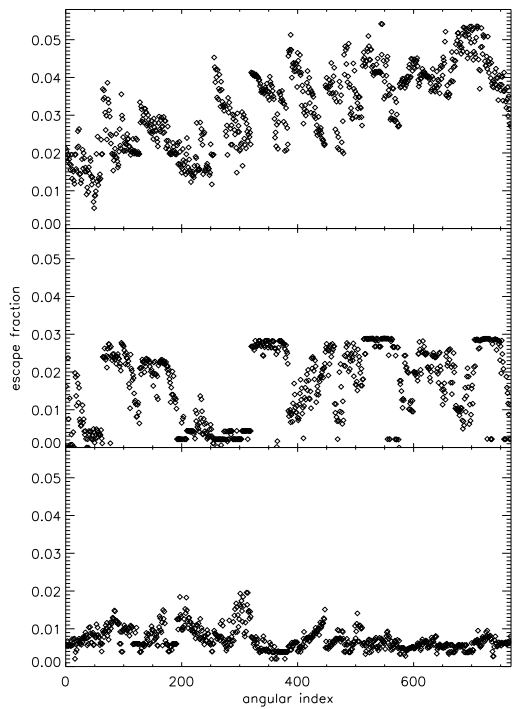}
  \caption{Same as Fig.~\ref{angularVariation.K15-06-8}, but for K93.}
  \label{angularVariation.K93-06-64}
\end{figure}

In this section we examine variations in {\it apparent} LyC
luminosities caused by orientation effects, i.e. that galaxies will be
oriented in a random way with respect to an observer. The
source-to-source scatter in $\fesc$ (Fig.~\ref{radial}) and a
non-uniform gas distribution in host galaxies invariably translate
into angular variations. In
Fig.~\ref{angularVariation.K15-06-8}--\ref{angularVariation.K93-06-64}
we plot the angular dependence of $\fesc$ for two galaxies, the more
massive 15 and the small 93, at three different redshifts, both
galaxies computed with the Kroupa IMF. For Salpeter galaxies the
behaviour is qualitatively the same. In both plots the x-axis gives
the index of the angular pixel in the Healpix \citep{gorski...02}
notation, covering the entire $4\pi$ in $768=12\times4^3$
bins. Essentially, it is as an index along a fractal space-filling
curve going through all bins on the sky (as seen from the galaxy) in
which $\fesc$ was computed, with each bin covering the same solid
angle. For the more massive galaxy 15, especially with the Kroupa IMF,
we can see two spikes corresponding to the face-on orientation of the
galaxy, with $\fesc$ $\sim2-3$ times the average.


\begin{figure}
  \epsscale{1.1}\plotone{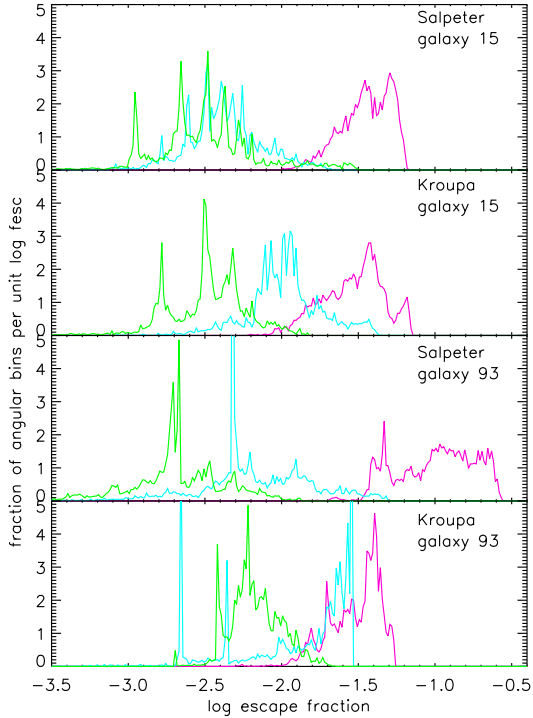}
  \caption{Probability function of the angular distribution of the
    Lyman-limit escape fraction for galaxies 15 and 93 at $z=3.6$
    (magenta), $z=2.95$ (cyan), and $z=2.39$ (green), with Salpeter
    and Kroupa IMFs as indicated in each panel.}
  \label{angularProbability}
\end{figure}

In Fig.~\ref{angularProbability} we plot the angular probability
distribution function of $\fesc$ for two Salpeter and two Kroupa
galaxies, at all three redshifts. The vertical axis shows the fraction
of angular bins per unit log $\fesc$, such that the area under each
curve is exactly unity. The decline of $\fesc$ with time is evident in
all four panels, but more interesting is the scatter of about an order
of magnitude in $\fesc$ depending on orientation to the
observer. This, together with the intrinsic LyC luminosity variation
from galaxy to galaxy (Fig.~\ref{luminosities-K98-S}) explains why in
magnitude-limited surveys only a fraction of galaxies are detected in
direct LyC emission \citep{shapley....06}.

\section{Conclusions}

In conclusion, we presented calculations of the LyC emission from 11
(proto-) disk galaxies in the redshift range $z=3.6-2.4$, modeled with
the standard Salpeter and Kroupa IMFs. To calculate the ionization
structure of the interstellar medium in each galaxy, we performed
time-dependent numerical radiative transfer around a large number (up
to $10^4$) of SF regions modeled with discrete star particles. Our
findings are as follows:

\smallskip

1) Although we find significant variations in the escape fraction of
ionizing photons from galaxy to galaxy, we confirm our earlier results
on the average decrease of $\fesc$ from $z=3.6$ to 2.4 for the entire
range of galaxy masses, as more gas cools and accretes onto galaxies,
forming stars in progressively clumpier environments. The resulting
LyC luminosity of individual galaxies declines gradually over this
redshift interval, by nearly an order of magnitude for lower-mass
galaxies, and a smaller factor for more massive objects. We also note
that this result is not sensitive to the assumed IMF, at least within
the range of feedback parameters normally associated with the widely
used Salpeter and Kroupa IMFs.

2) The observed galaxy-to-galaxy scatter in LyC emission is caused in
approximately equal parts by the inclination effects and the intrinsic
variations in the $4\pi$-averaged UV luminosities, for galaxies of a
fixed mass.

3) We predict that the cosmic galactic ionizing UV luminosity would be
monotonically decreasing from $z=3.6$ to 2.4. Some fraction of smaller
galaxies in clusters can turn on their SF for the first time during
this redshift interval due to tidal effects, however, the intrinsic
changes in most field galaxies associated with continuous gas infall
would lead to a decline in the LyC comoving luminosity density.

On the other hand, there is evidence that the galaxy non-ionizing UV
luminosity function gradually rises at the faint end from $z=4$ to
$z=2$ \citep{sawicki.06}. We want to stress that our findings do not
contradict these data, as indeed we see a rise by a factor of
$\sim2-4$ in the number of young stars from $z=3.6$ to 2.4 in all our
small galaxies, consistent with continuous gas infall. We are planning
to study the non-ionizing UV luminosity function in the same set of
galaxies in the future.

4) The higher escape fractions as well as LyC luminosities found at
$z=3.6$, compared to the lower redshifts, also support the notion that
galaxies become a progressively more important source of ionizing
photons as one goes back in time, as the comoving number density of
quasars declines rapidly at $z\gsim3$ \citep{richards-06}.

\acknowledgments

We wish to thank Akio Inoue for drawing our attention to the question
of dispersion of observed $\fesc$ along various lines of sight, and
Marcin Sawicki for careful reading of the manuscript. We would also
like to thank the referee for very useful comments. Numerical
calculations were done on the SGI Itanium II facility provided by the
Danish Centre for Scientific Computing and on the Linux cluster of the
Institute for Computational Astrophysics. The Dark Cosmology Centre is
funded by the DNRF.

\bibliographystyle{apj}

\begin{thebibliography}{20}
\expandafter\ifx\csname natexlab\endcsname\relax\def\natexlab#1{#1}\fi

\bibitem[{{Abel} \& {Wandelt}(2002)}]{abel.02}
{Abel}, T. \& {Wandelt}, B.~D. 2002, MNRAS, 330, L53

\bibitem[{{Dahlen} {et~al.}(2007){Dahlen}, {Mobasher}, {Dickinson}, {Ferguson},
  {Giavalisco}, {Kretchmer}, \& {Ravindranath}}]{dahlen......07}
{Dahlen}, T., {Mobasher}, B., {Dickinson}, M., {Ferguson}, H.~C., {Giavalisco},
  M., {Kretchmer}, C., \& {Ravindranath}, S. 2007, \apj, 654, 172

\bibitem[{{G{\'o}rski} {et~al.}(2002){G{\'o}rski}, {Banday}, {Hivon}, \&
  {Wandelt}}]{gorski...02}
{G{\'o}rski}, K.~M., {Banday}, A.~J., {Hivon}, E., \& {Wandelt}, B.~D. 2002, in
  ASP Conf. Ser. 281: Astronomical Data Analysis Software and Systems XI, ed.
  D.~A. {Bohlender}, D.~{Durand}, \& T.~H. {Handley}, 107

\bibitem[{{Greve} \& {Sommer-Larsen}(2006)}]{greve.06}
{Greve}, T.~R. \& {Sommer-Larsen}, J. 2006, ArXiv Astrophysics e-prints

\bibitem[{{Inoue} {et~al.}(2006){Inoue}, {Iwata}, \& {Deharveng}}]{inoue..06}
{Inoue}, A.~K., {Iwata}, I., \& {Deharveng}, J.-M. 2006, MNRAS, 371, L1

\bibitem[{{Kroupa}(1998)}]{kroupa98}
{Kroupa}, P. 1998, MNRAS, 298, 231

\bibitem[{{Laursen} \& {Sommer-Larsen}(2007)}]{laursen.07}
{Laursen}, P. \& {Sommer-Larsen}, J. 2007, \apjl, 657, L69

\bibitem[{{Leitherer} {et~al.}(1999){Leitherer}, {Schaerer}, {Goldader},
  {Delgado}, {Robert}, {Kune}, {de Mello}, {Devost}, \&
  {Heckman}}]{leitherer........99}
{Leitherer}, C., {Schaerer}, D., {Goldader}, J.~D., {Delgado}, R.~M.~G.,
  {Robert}, C., {Kune}, D.~F., {de Mello}, D.~F., {Devost}, D., \& {Heckman},
  T.~M. 1999, ApJS, 123, 3

\bibitem[{{Lia} {et~al.}(2002){Lia}, {Portinari}, \& {Carraro}}]{lia..2002}
{Lia}, C., {Portinari}, L., \& {Carraro}, G. 2002, \mnras, 330, 821,
erratum \mnras, 335, 864

\bibitem[{{Mac Low} \& {Ferrara}(1999)}]{maclow.99}
{Mac Low}, M.-M. \& {Ferrara}, A. 1999, \apj, 513, 142

\bibitem[{{Razoumov} \& {Sommer-Larsen}(2006)}]{razoumov.06}
{Razoumov}, A.~O. \& {Sommer-Larsen}, J. 2006, ApJ, 651, L89

\bibitem[{{Richards} {et~al.}(2006){Richards}, {Strauss}, {Fan}, {Hall},
  {Jester}, {Schneider}, {Vanden Berk}, {Stoughton}, {Anderson}, {Brunner},
  {Gray}, {Gunn}, {Ivezi{\'c}}, {Kirkland}, {Knapp}, {Loveday}, {Meiksin},
  {Pope}, {Szalay}, {Thakar}, {Yanny}, {York}, {Barentine}, {Brewington},
  {Brinkmann}, {Fukugita}, {Harvanek}, {Kent}, {Kleinman}, {Krzesi{\'n}ski},
  {Long}, {Lupton}, {Nash}, {Neilsen}, {Nitta}, {Schlegel}, \&
  {Snedden}}]{richards-06}
{Richards}, G.~T., {Strauss}, M.~A., {Fan}, X., {Hall}, P.~B., {Jester}, S.,
  {Schneider}, D.~P., {Vanden Berk}, D.~E., {Stoughton}, C., {Anderson}, S.~F.,
  {Brunner}, R.~J., {Gray}, J., {Gunn}, J.~E., {Ivezi{\'c}}, {\v Z}.,
  {Kirkland}, M.~K., {Knapp}, G.~R., {Loveday}, J., {Meiksin}, A., {Pope}, A.,
  {Szalay}, A.~S., {Thakar}, A.~R., {Yanny}, B., {York}, D.~G., {Barentine},
  J.~C., {Brewington}, H.~J., {Brinkmann}, J., {Fukugita}, M., {Harvanek}, M.,
  {Kent}, S.~M., {Kleinman}, S.~J., {Krzesi{\'n}ski}, J., {Long}, D.~C.,
  {Lupton}, R.~H., {Nash}, T., {Neilsen}, Jr., E.~H., {Nitta}, A., {Schlegel},
  D.~J., \& {Snedden}, S.~A. 2006, \aj, 131, 2766

\bibitem[{{Salpeter}(1955)}]{salpeter55}
{Salpeter}, E.~E. 1955, ApJ, 121, 161

\bibitem[{{Sawicki} \& {Thompson}(2006)}]{sawicki.06}
{Sawicki}, M. \& {Thompson}, D. 2006, \apj, 642, 653

\bibitem[{{Shapley} {et~al.}(2006){Shapley}, {Steidel}, {Pettini},
  {Adelberger}, \& {Erb}}]{shapley....06}
{Shapley}, A.~E., {Steidel}, C.~C., {Pettini}, M., {Adelberger}, K.~L., \&
  {Erb}, D.~K. 2006, ApJ, 651, 688

\bibitem[{{Sommer-Larsen}(2006)}]{sommer-larsen06}
{Sommer-Larsen}, J. 2006, ApJ, 644, L1

\bibitem[{{Sommer-Larsen} \& {Dolgov}(2001)}]{sommer-larsen.01}
{Sommer-Larsen}, J. \& {Dolgov}, A. 2001, \apj, 551, 608

\bibitem[{{Sommer-Larsen} {et~al.}(2003){Sommer-Larsen}, {G{\"o}tz}, \&
  {Portinari}}]{sommer-larsen..03}
{Sommer-Larsen}, J., {G{\"o}tz}, M., \& {Portinari}, L. 2003, ApJ, 596, 47

\bibitem[{{Springel} \& {Hernquist}(2002)}]{springel.02}
{Springel}, V. \& {Hernquist}, L. 2002, \mnras, 333, 649

\bibitem[{{Trujillo} {et~al.}(2006){Trujillo}, {F{\"o}rster Schreiber},
  {Rudnick}, {Barden}, {Franx}, {Rix}, {Caldwell}, {McIntosh}, {Toft},
  {H{\"a}ussler}, {Zirm}, {van Dokkum}, {Labb{\'e}}, {Moorwood},
  {R{\"o}ttgering}, {van der Wel}, {van der Werf}, \& {van
  Starkenburg}}]{trujillo.06}
{Trujillo}, I., {F{\"o}rster Schreiber}, N.~M., {Rudnick}, G., {Barden}, M.,
  {Franx}, M., {Rix}, H.-W., {Caldwell}, J.~A.~R., {McIntosh}, D.~H., {Toft},
  S., {H{\"a}ussler}, B., {Zirm}, A., {van Dokkum}, P.~G., {Labb{\'e}}, I.,
  {Moorwood}, A., {R{\"o}ttgering}, H., {van der Wel}, A., {van der Werf}, P.,
  \& {van Starkenburg}, L. 2006, \apj, 650, 18

\end{thebibliography}

\end{document}